\documentclass[twocolumn,aps,amsmath,amssymb,showpacs]{revtex4}

\usepackage{graphicx}% Include figure files
\usepackage{dcolumn}% Align table columns on decimal point
\usepackage{bm}% bold math

\begin{document}
\title{Observation of shell effects in nanowires for the noble metals copper, silver and gold}

\author{A.I. Mares and J.M. van Ruitenbeek}
\affiliation{Kamerlingh Onnes Laboratorium, Universiteit Leiden,
P.O. Box 9504, 2300 RA Leiden, The Netherlands}

\date{\today}

\begin{abstract}
We extend our previous shell effect observation in gold nanowires
at  room temperature under ultra high vacuum to the other two
noble metals: silver and copper. Similar to gold, silver nanowires
present two series of exceptionally stable diameters related to
electronic and atomic shell filling. This observation is in
concordance to what was previously found for alkali metal
nanowires. Copper however presents only electronic shell filling.
Remarkably we find that shell structure survives under ambient
conditions for gold and silver.
\end{abstract}


\maketitle

\section{Introduction}

Evidence shows that the stability of metallic nanowires is
strongly correlated to their electrical properties. Applying a
free electron model to a cylindrical nanowire, the electronic free
energy as a function of the radius shows an oscillating spectrum
with minima that represent stable nanowire configurations due to
shell filling \cite{stafford97}. Experimental evidence of shell
filling in metallic nanowires was reported for alkali metal
nanowires by Yanson \textit{et al}. \cite{yanson99}. Similar to
metal clusters \cite{deheer93}, alkali metal nanowires present two
series of stable diameters, due to electronic and atomic shell
filling \cite{yanson01a}.\\
In our previous work \cite{mares04} we reported evidence that
shell filling effects are also present in gold nanowires. In this
paper we extend the study to the other two monovalent noble
metals: silver and copper. The noble metal nanowires are more
suitable for applications, being less reactive than the alkali
metal nanowires. It would be of great importance to be able to
predict and control nanowire stability. Noble metals differ from
alkali ones in the shape of Fermi surface (nearly spherical vs
almost perfectly spherical) and also in the bulk packing (fcc vs
bcc). To some extend the free electron model can be applied also
to noble metals nanowires, as was proven successfully for noble
metal clusters \cite{deheer93}. We present evidence that, similar
to gold, silver and copper nanowires show certain exceptionally
stable diameters of the same origin: shell filling. Firstly, we
see electronic shell effects in all three metals. Secondly, the
atomic shell effect appears only in gold and silver nanowires.
Silver however, is exceptional, regarding the more pronounced
shell structure as well as the small variation in the peak
positions. Remarkably, we find that for gold and silver some of
the stable diameters survive even under ambient conditions, which
is a big step in the direction of possible applications.
\begin{figure}[b!]
\begin{center}
\includegraphics[width=7cm]{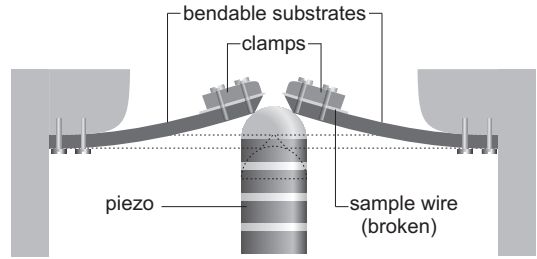}
\end{center}
\noindent{\caption{Schematic view of the MCBJ technique under UHV.
The notched sample wire is clamped onto two separate beams and
broken by bending the beams. Contact between the fracture surfaces
can be finely adjusted by means of the piezoelectric element. The
relaxed configuration of the sample is represented by the dotted
lines.
 } \label{bendratioPRL}}
\end{figure}

\section{Experimental technique}

The stability analysis of the noble metal nanowires is done by
investigating electrical conductance using a mechanically
controllable break junction  (MCBJ) method. A bulk
poly-crystalline metal wire is notched circularly and fixed on a
substrate. By bending the substrate with a piezoelectric element
the wire breaks at the most sensitive point, the notch. By
retracting the piezoelement the contact between the two bulk
pieces will be remade. Controlling the voltage on the
piezoelectric element, one can finely control the dimensions of
the contact with atomic resolution. In the process of thinning
down, the contact experiences different metastable configurations,
depending on the atomic rearrangements in the nanowire and its
close vicinity.\\
Since we search for stable diameters, the atoms need to have
sufficient mobility to select the most favorable among all
possible metastable configurations. One way to enhance their
mobility is by increasing the thermal energy. The optimal
temperature is a significant fraction of the melting temperature
but one has to take into account that for nanowires the melting
temperature is strongly suppressed. For example Hwang \textit{et
al}. \cite{kwang03} find in a calculation for copper nanowires of
34 atoms in cross section a meting temperature of 590 K (compared
to bulk value 1357 K). On the high end the optimal temperature is
limited by the reduced lifetime of the metastable states at
elevated temperatures. B{\" u}rki \textit{et al}. \cite{burki05}
give an estimate of the relevant activation energies, which are
more than a factor of two higher for the noble metals as compared
to the alkali metals.\\
We have developed a new MCBJ technique adapted to the use at
elevated temperatures in ultra high vacuum (UHV). The bending beam
consists of two bendable phosphor bronze substrates where the
notched wire is fixed by stainless steel clamps
(Fig.~\ref{bendratioPRL}). The bending beams are mounted on
separate stainless steel supports such that the only electrical
contact between them is formed by the wire
(Fig.~\ref{bendratioPRL}). The base pressure during measurements
is $4\times10^{-10}$ mbar and the temperature range achievable is
approximately $70$K- $500$K. The cooling down is done by thermal
contact of the sample substrate with a nitrogen bath. The sample
can also be locally heated by radiation from a tungsten filament.
The temperature is monitored by a type E thermocouple. We have
improved our previous design such that our new sample holder has a
tray of six bending beams with a sample mounted on each, that we
can independently measure, avoiding in this way to break the
vacuum for each new wire.\\
The conductance is measured at constant bias voltage, recording
the current with a current-voltage converter using a digital to
analog card of 16 bits resolution. The contact is thinned down
starting from about 100 $G_{0}$ in about one second. Here,
$G_{0}=2e^{2}/h$ is the quantum of conductance and 100 $G_{0}$
roughly corresponds to $\sim$ 100 atoms in cross section.
Different breaking times in the range of 10 ms to few minutes and
different dimensions of the starting contact were tested. A
typical conductance trace follows a step like pattern, with
plateaus for metastable configurations of the contact and jumps
resulting from atomic rearrangements in the vicinity of the
contact. In order to find only the preferred diameters from all
metastable configurations, we use a statistical analysis, by
adding many conductance traces in a histogram. The conductance
scale is divided into about 600 bins, and a histogram is build
from a few thousands of scans. A peak in the histogram corresponds
to a preferred, reproducible configuration of the contact.\\
\section{Results}
\subsection{Electronic shell effects}
\begin{figure}[t!]
\begin{center}
\includegraphics[width=8cm]{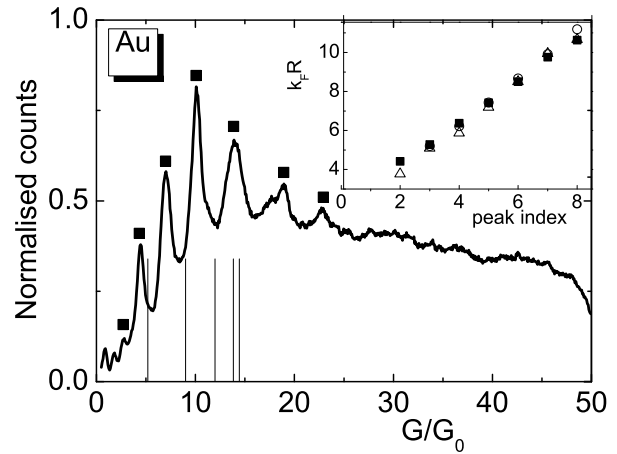}
\end{center}
\caption{Conductance histogram for gold at room temperature under
UHV constructed from 5000 individual consecutive traces, using a
bin size of 0.1\,$G_0$ and a bias voltage of 100\,mV, giving
evidence of electronic shell filling. Peak positions are indicated
by squares.  By bars we plot the calculated conductance for
helical nanowires \protect\cite{Ono05}. The inset shows the peak
positions, converted to $k_{\rm F}R$ with the help of
Eq.~(\protect\ref{Sharvin}), as a function of peak index (filled
squares), magic radii for gold clusters (circles)
\protect\cite{katakuse85}, and predictions of the minima of the
electronic energy calculation (triangles) taken from
\protect\cite{Ogando02}. The experimentally observed periodicity
of the peaks is $\Delta k_{\rm F}R= 1.06 \pm 0.01$.}
\label{GAuESE}
\end{figure}

Fig. \ref{GAuESE} presents a conductance histogram recorded at
room temperature under UHV (UHV-RT) using a bias voltage of 100
mV. The histogram reproduces our previously reported result
\cite{mares04}. One can see a sequence of distinct peaks at
certain conductance values. In the low conductance range peaks are
situated close to 1, 2 ,3 $G_{0}$ the conductance for 1, 2, 3
atoms in cross section, as reported previously for gold atomic
contacts \cite{costa97,yanson01t}. We see that these peaks have a
relatively low amplitude, the maximum being at the peak of 10
$G_{0}$.\\
For the regime of thick nanowires the conductance is related to
the nanowire radius by a semi-classical formula for a ballistic
nanowire with circular cross section:
\begin{equation}\label{Sharvin}
  G=g G_{0}\cong G_{0}\left[\left(\frac{k_{\rm F}R}{2}\right)^{2}-\frac{k_{\rm F}R}{2}+\frac{1}{6}+...\right],
\end{equation}
with $k_{\rm F}R$ the Fermi wave vector, g the reduced conductance
and R the radius of the nanowire \cite{torres94},
\cite{hoppler98}. When we plot the peak positions in units $k_{\rm
F}R$ as function of peak index we get a linear dependence,
illustrated in the inset of Fig. \ref{GAuESE} with a slope $\Delta
k_{\rm F}R=1.06\pm 0.01$, similar to the one obtained for alkali
metals. This is an indication that the peaks in the conductance
histogram are due to electronic shell filling: the nanowire
chooses such diameters that
give minima in the electronic free energy.\\
\begin{figure}[t!]
\begin{center}
\includegraphics[width=8cm]{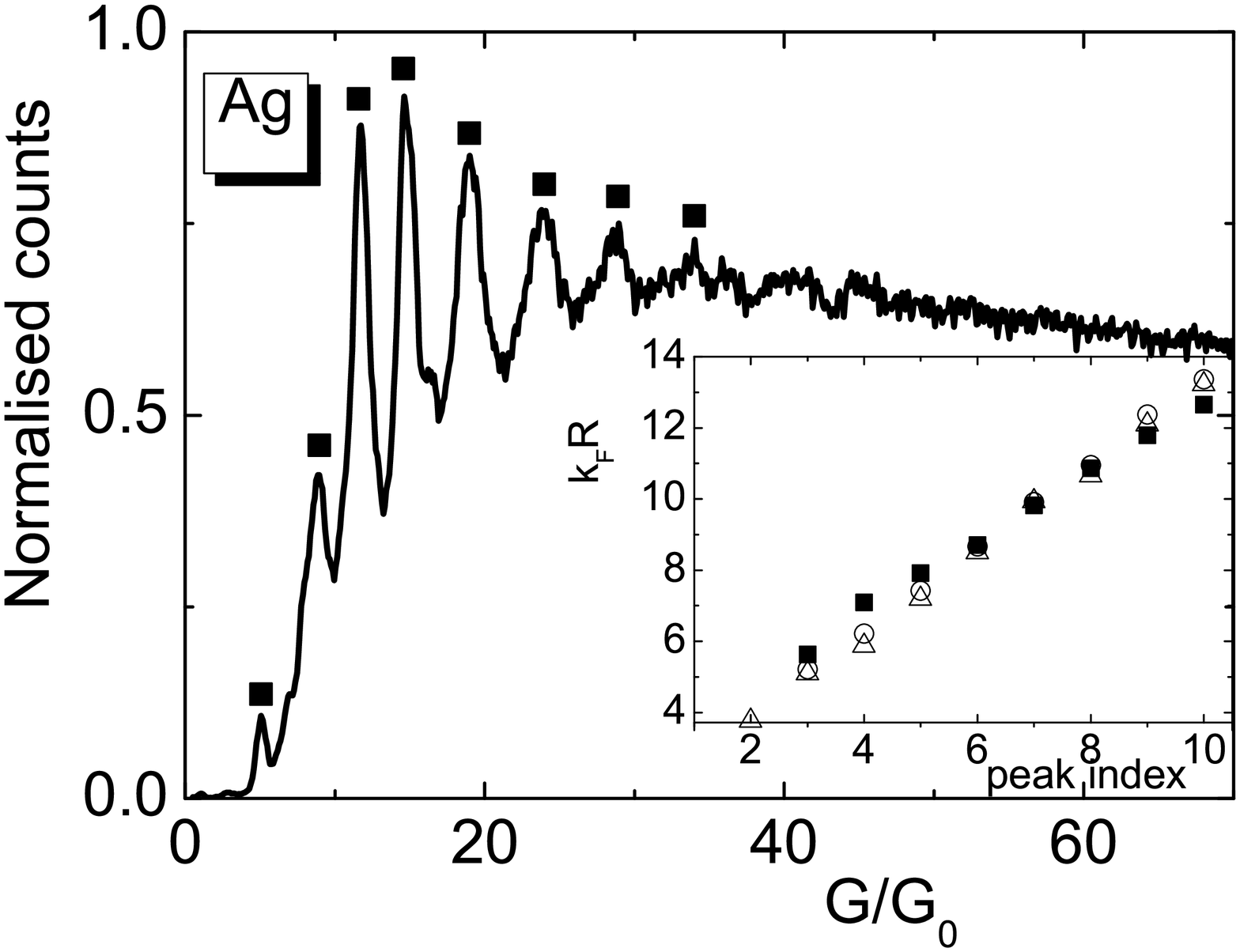}
\includegraphics[width=8cm]{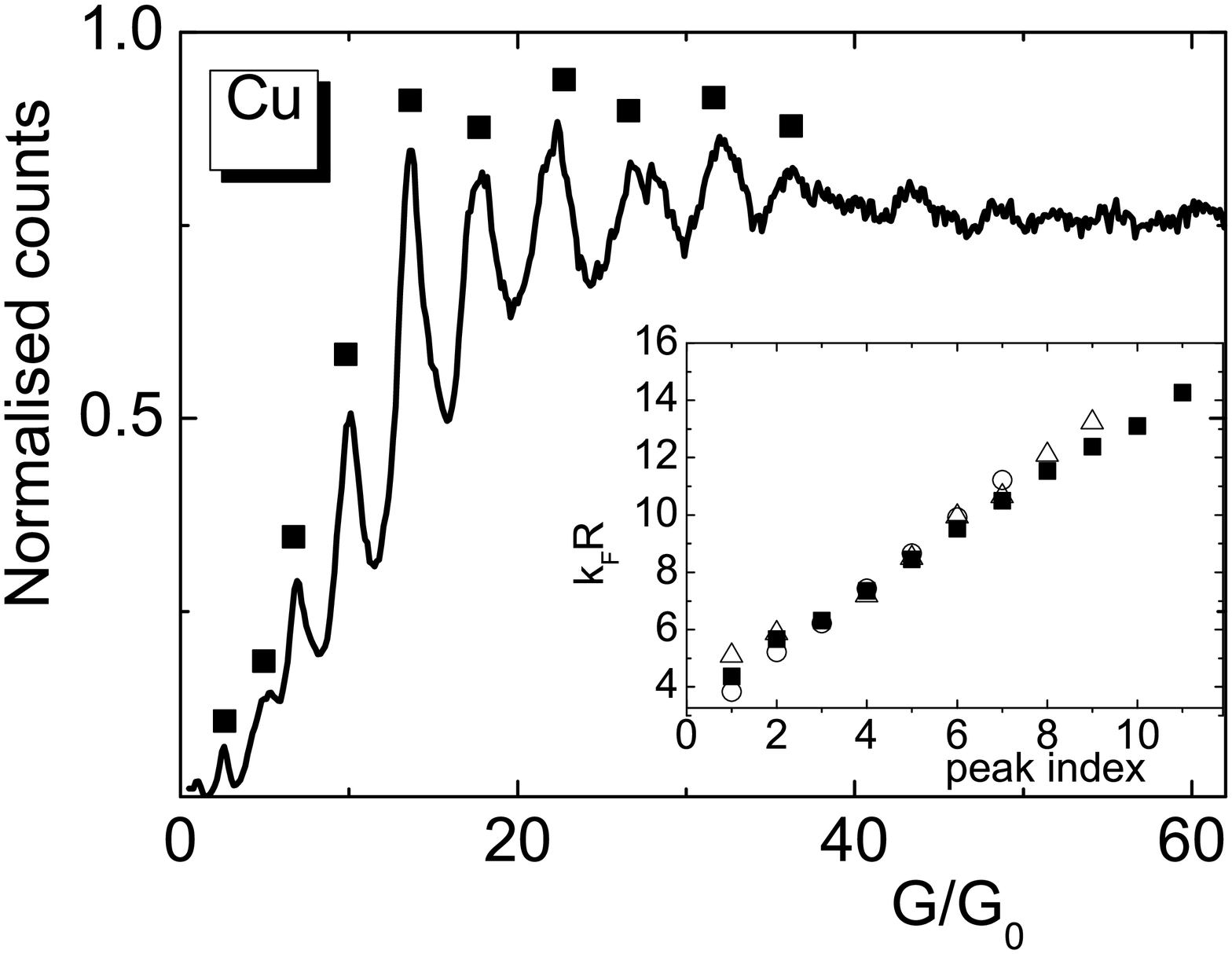}
\end{center}
\caption{Conductance histogram for silver (top) and copper
(bottom) at room temperature, giving evidence of electronic shell
filling. The silver histogram is constructed from 10000 individual
consecutive traces, using a bin-size of 0.1\,$G_0$, while for
copper 20000 individual consecutive traces were included and a
bin-size of 0.14\,$G_0$ was used. In each case the bias voltage
was 100\,mV. The insets show the peak positions, converted to
$k_{\rm F}R$, as a function of peak index (filled squares). The
slope is $\Delta k_{\rm F}R=0.98\pm 0.01$, both for silver and
copper. Magic radii for silver and copper clusters
\cite{katakuse85} and theoretical predictions for stable diameters
in nanowires \cite{Ogando02} are shown for comparison (circles and
triangles, respectively).} \label{GAgandCuESE}
\end{figure}
%
%\begin{figure}[t!]
%\begin{center}
%\includegraphics[width=8cm]{GCuESE-4.eps}
%\end{center}
%\caption{Conductance histogram for copper at room temperature
%constructed from 20000 individual consecutive traces, using a
%bin-size of 0.14\,$G_0$ and a bias voltage of 100\,mV. The
%oscillations in the histogram constitute an evidence of electron
%shell filling, as illustrated by the linear dependence of the peak
%positions, converted to $k_{\rm F}R$, as a function of peak index
%(inset, filled squares) with a slope $\Delta k_{\rm F}R=0.98\pm
%0.01$. Magic radii in copper clusters taken from
%\cite{katakuse85}(circles) and theoretical prediction of free
%electron model taken from \cite{Ogando02}(triangles) are presented
%for comparison.} \label{GCuESE}
%\end{figure}
%
We now find similar periodic patterns for silver and copper
nanowires as one can see in the histograms of
Fig.~\ref{GAgandCuESE}. The periodicity of the peaks is similar to
gold. Thus for silver and copper the slope is $\Delta k_{\rm
F}R=0.98\pm 0.01$. The maximum spectrum amplitude for silver is
found at about 15 $G_{0}$, while for gold and copper it varies
between different measurements on values 7, 10, 12 for gold and
10, 14, 18 for
copper.\\
\begin{figure}[t!]
\begin{center}
\vskip 0.2cm
\includegraphics[width=8cm]{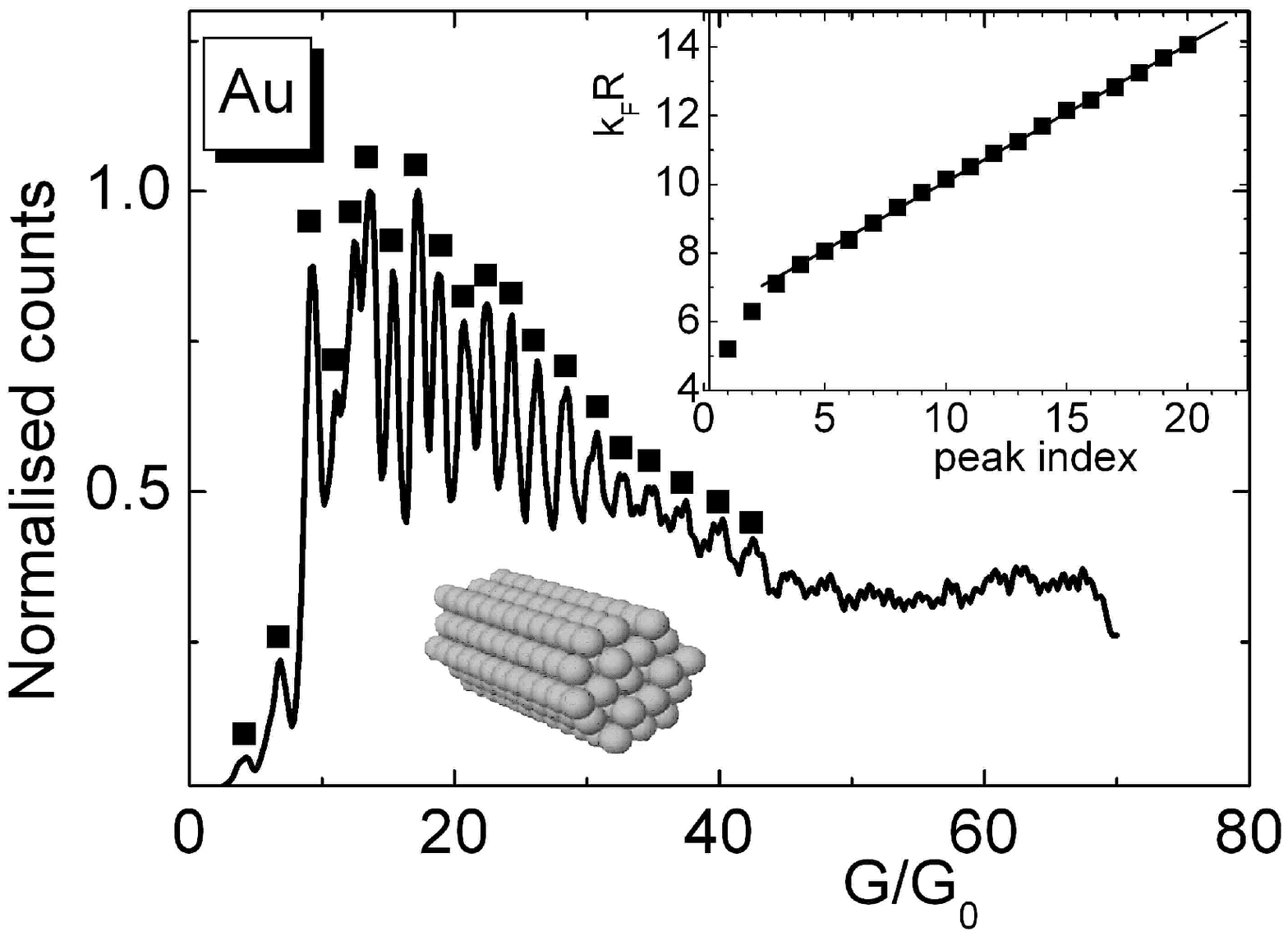}
\includegraphics[width=8cm]{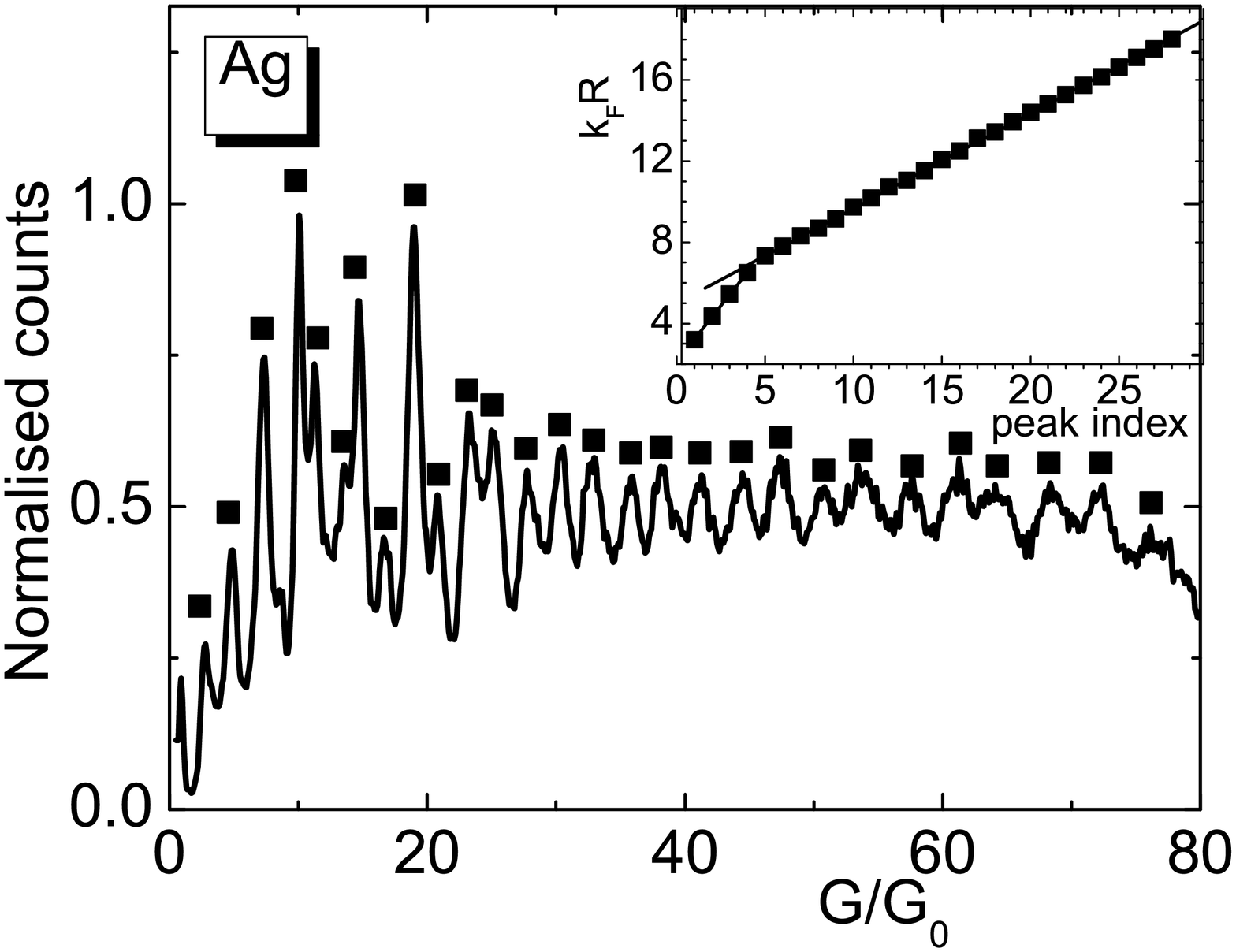}
\end{center}
\noindent{\caption{Conductance histogram for Au (top) and silver
(bottom) obtained from 3000 and 4500 individual conductance
traces, respectively, recorded under UHV-RT, giving evidence of
electronic and atomic shell effects. The bias voltage was 150 mV
for gold and 100 mV for the silver measurements. We observe a
cross over from electronic to atomic shell structure at G$\sim$ 10
G$_0$. Peak positions as function of peak index (top insets)
exhibit a linear dependence as expected for atomic shell effect
with slopes of $\Delta k_{\rm F}R=0.400\pm 0.002$ and $\Delta
k_{\rm F}R=0.460\pm 0.001$ for gold and silver, respectively. The
lower inset in the top panel shows a sketch of a nanowire along
the [110] axis with hexagonal cross section with four (111) facets
and two larger (100) ones.}\label{GAuandAgASE}}
\end{figure}
%
%\begin{figure}[t!]
%\begin{center}
%\includegraphics[width=8cm]{GAgASE-4.eps}
%\end{center}
%\noindent{\caption{UHV-RT conductance histogram for Ag obtained
%from 4500 individual conductance traces, recorded at a bias
%voltage of 100 mV. The crossover between electronic and atomic
%shell effect is at G$\sim$ 10 G$_0$. The period of the atomic
%shell effect peaks determined from the slope of peaks position as
%function of peak index (inset) is $\Delta k_{\rm F}R=0.46\pm
%0.001$. }\label{GAgASE}}
%\end{figure}
%
%
\subsection{Atomic shell effects}

Sometimes a new series of peaks appears in the histogram as we can
see in Fig.~\ref{GAuandAgASE} (top), that was reported in our
previous work \cite{mares04} recorded for gold in UHV-RT. This is
related to a geometrical effect also present in clusters, namely,
atomic shell filling. Certain nanowires are more stable when they
adopt a crystalline order with smooth facets such as to obey
minima of surface energy. This effect is expected to appear at
larger diameters than electronic shell filling. Silver nanowires
present this new series of stable diameters even more pronounced
than gold does , with peaks up to conductance values of 80 $G_{0}$
(see Fig.~\ref{GAuandAgASE}). However, for copper we have not
observed distinct atomic shell effect peaks.\\
The crossover between electronic and atomic shell effects is in
most of the cases around 10 $G_{0}$ for gold, and at about 15
$G_{0}$ for silver but it can vary around this value between
different measurements. This variation can be due to local
crystalline orientation, a parameter that we cannot control during
measurements. The crossover value is in some histograms hard to
determine since besides the consecutive series of peaks having
atomic shell effect periodicity electronic shell effect peaks
appear to be superimposed.\\
\begin{figure}[t!]
\begin{center}
\includegraphics[width=8cm]{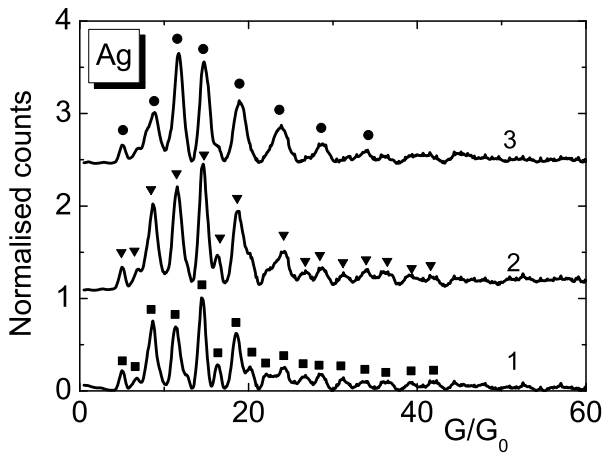}
\end{center}
\noindent{\caption{Evolution of Ag conductance histograms obtained
in UHV-RT recorded at a bias voltage of 100 mV during the same
measurement, containing 5000 traces (1), 8000 traces (2), and
20000 traces (3). Histogram (3) includes the traces of histograms
(2) and (1). From all three curves a smooth background was
subtracted. In curves (1) and (2) the peaks obey atomic shell
effect period, while in curve (3) a transition to electronic shell
effect period occurs. }\label{Gaseese}}
\end{figure}
We observe that during a particular measurement, after repeated
cycles of making and breaking the contact, an evolution from
atomic shell effect to electronic shell effect appears, as one can
see in Fig. \ref{Gaseese}. Curves 1, 2, 3 are histograms recorded
during the same measurement containing 5000, 8000, and 20000
consecutive scans. A smooth positive background was subtracted
from the histograms for better clarity. Firstly we can see that
some peaks having atomic shell effect periodicity in histogram 1
gradually decrease their weight in histogram 2 until they
disappear in histogram 3 (peaks at G $\sim$ 7, 16, 20, 22, 26, and
all the peaks above this value). Secondly we see that in histogram
3 the peaks vanish above 30 G$_0$, while in histograms 1, 2 they
are visible up to about 40 G$_0$. Finally in the histogram 3 we
get peaks that have electronic shell effect periodicity. This
transition from atomic to electronic shell effect was reported
previously also for alkali metals \cite{yanson01}, and can be due
to an increase in mobility of the atoms during repeated cycles of
elongation/compression of the nanowire, that can damage the
faceting. Another possible reason may be that during repeated
indentation the crystalline orientation of the nanowire or of the
connecting electrodes changes, not being favorable anymore for
faceting.

\subsection{Experiments under ambient conditions}

\begin{figure}[t!]
\vskip 0.2cm
\begin{center}
\includegraphics[width=8cm]{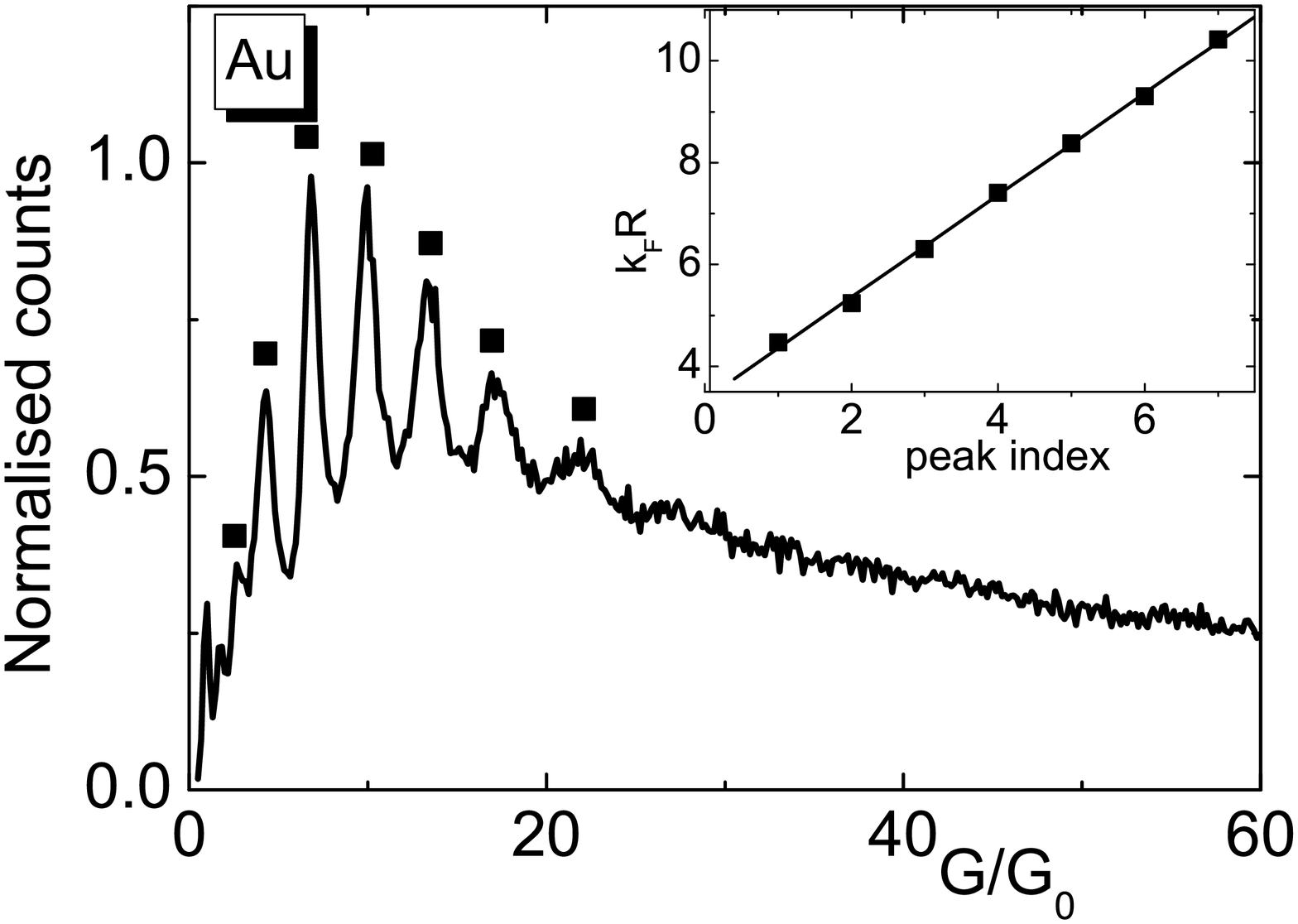}
\includegraphics[width=8cm]{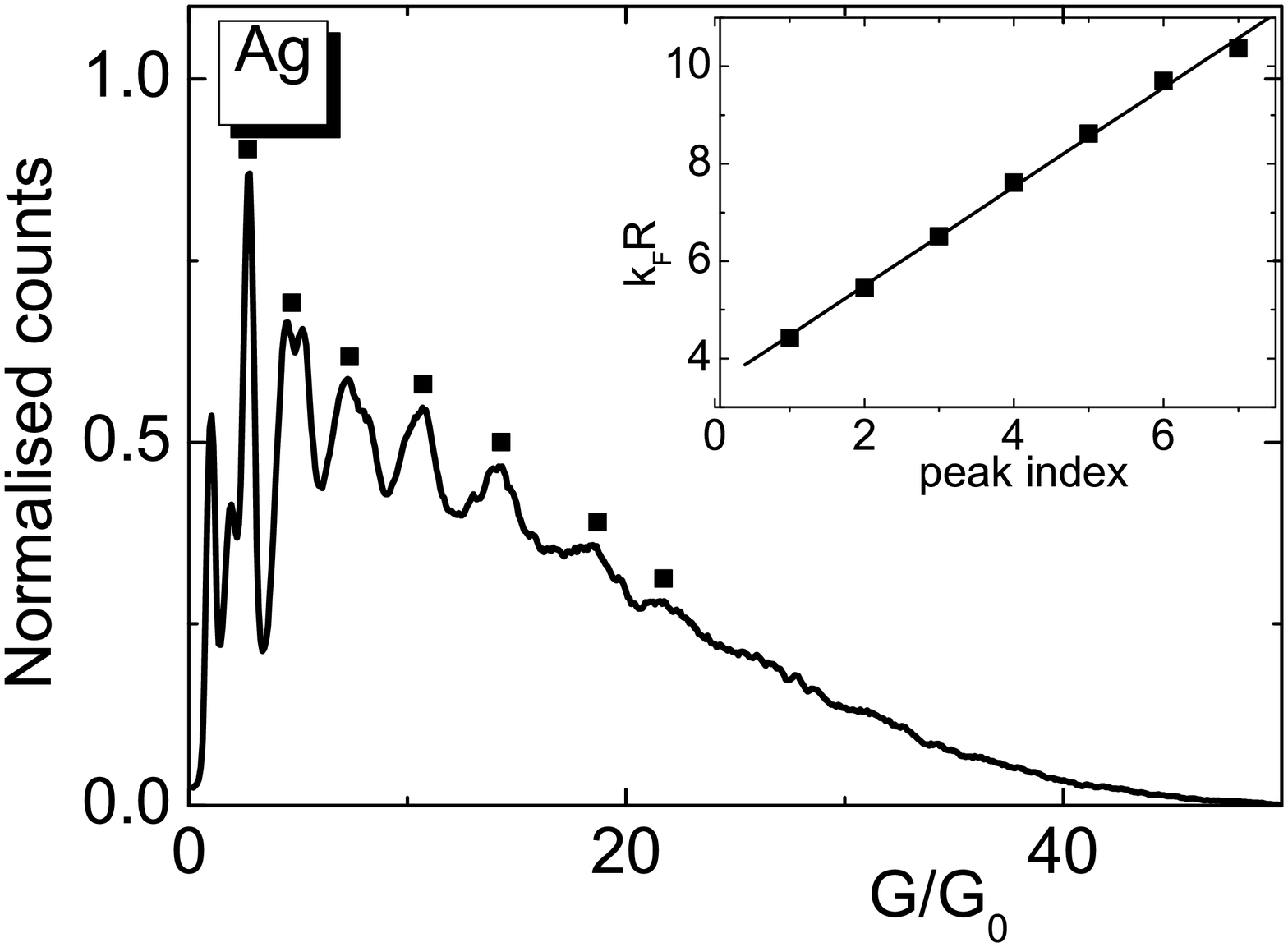}
\end{center}
\caption{Conductance histograms for gold (top) and silver (bottom)
at room temperature under ambient conditions, constructed from
2000 and 10000 individual consecutive traces, respectively, giving
evidence of electronic shell filling. The bin size was 0.13\,$G_0$
(gold) and 0.08\,$G_0$ (silver) and the bias voltage of 100\,mV
(gold) and 20\,mV (silver). The insets show the peak positions,
converted to $k_{\rm F}R$, as a function of peak index with a
slope $\Delta k_{\rm F}R=1.00 \pm 0.01$ (gold) and $\Delta k_{\rm
F}R=1.06 \pm 0.02$ (silver).} \label{GAuandAgAir}
\end{figure}
Fig.~\ref{GAuandAgAir} (top) shows a conductance histogram for
gold recorded at room temperature under ambient conditions. One
can clearly distinguish peaks up to about 22 $G_{0}$, with a
periodicity $\Delta k_{\rm F}R=1.00\pm 0.01$, very close to the
value obtained in UHV. The peak positions are close to the ones
obtained in UHV, although some of them may be shifted somewhat to
lower values. Similarly, a silver conductance histogram in air
shows electronic shell effect periodicity $\Delta k_{\rm
F}R=1.06\pm 0.02$ (Fig.~\ref{GAuandAgAir}, bottom). This brings
evidence that, remarkably, shell structure survives even under
ambient conditions in silver and gold. The relative intensity of
the peaks is different from those under UHV. The maximum amplitude
is shifted to lower conductance values with respect to UHV.
Moreover the peak at about 1 $G_{0}$, commonly attributed to a
single-atom contact \cite{scheer98} has a much higher amplitude
than under UHV. It has been shown by Hansen \textit{et al}.
\cite{hansen00} that a one-atom contact is hardly stable under
UHV-RT, due to the high mobility of the atoms. However, under
ambient conditions adsorbates decrease the atom mobility resulting
in an enhanced stability of small contacts. This may explain the
experiments on gold atomic contacts at RT in air \cite{costa97}.
In our conductance histograms  we see that only the electronic
shell effect survives in air. This is not unexpected since the
atomic shell effect is a surface effect, therefore adsorbed
species modify the surface energy and are expected to damage
the faceting.\\
Copper does not show shell effect peaks in air. The dominant
feature is a broad peak close to 1 $G_{0}$, as previously reported
\cite{hansen97}. Since copper is known to be the most reactive of
the three noble metals, the absence of shell structure can be
caused by fast oxidation of the contact.
%
%\begin{figure}[t!]
%\begin{center}
%\includegraphics[width=8cm]{GAgAir-4.eps}
%\end{center}
%\caption{Conductance histogram for silver at room temperature
%under ambient conditions, constructed from 10000 individual
%consecutive traces, using a bin-size of 0.08\,$G_0$ and a bias
%voltage of 20\,mV, giving evidence of electronic shell filling.
%The inset shows the peak positions, converted to $k_{\rm F}R$, as
%a function of peak index with a slope $\Delta k_{\rm F}R=1.06 \pm
%0.02$.} \label{GAgAir}
%\end{figure}

\section{Discussion}

\subsection{Comparison with low temperature histograms}

Conductance histograms for gold at low temperatures reported in
the literature typically show only the range of low conductances
that is dominated by a peak near 1 $G_{0}$,  attributed to a one
atom contact, see e.g., results on gold at liquid helium
temperatures \cite{yanson01t}. Peaks can be distinguished only up
to $3 G_{0}$ followed by a flat tail. For copper and silver
conductance histograms recorded at helium temperature are similar
to gold having a dominant peak at or just below $1G_{0}$, followed
by two additional peaks of lower intensity \cite{krans93}. There
is a major difference in the origin of the low temperature peaks
compared to our UHV-RT histograms. At low temperature the atoms
are frozen in configurations that have a certain conductance
value. In UHV-RT measurements atomic mobility plays an important
role and the nanowire can self-organize such to find the most
stable configuration. Therefore, the peaks in our data reflect
preferred stable diameters, and not preferred conductance like in
the case of low temperature histograms.

\subsection{Comparison between the three different noble metals}

\begin{figure}[b!]
\begin{center}
\includegraphics[width=8cm]{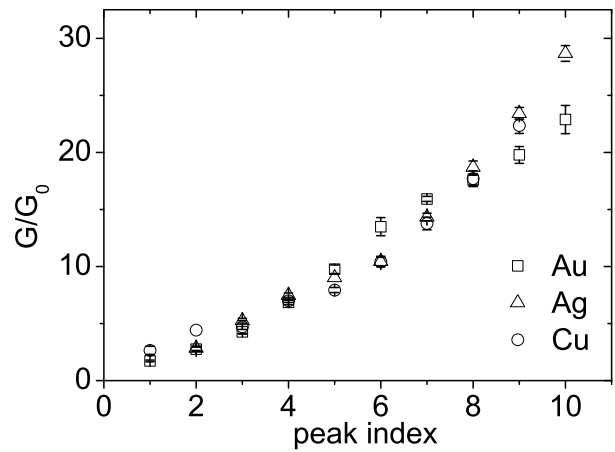}
\end{center}
\caption{Averaged peak positions and their standard deviations
obtained from conductance histograms of independent measurements
for Ag (11 measurements; squares), Cu (12 measurements; circles),
and Au (5 measurements; triangles). Results of the stabilized
jellium model considering a circular cross section are also
included (crosses) \protect\cite{Ogando02}. } \label{GpeaksAgCuAu}
\end{figure}

In Fig. \ref{GpeaksAgCuAu} we plot the averaged values of the peak
positions in histograms showing electronic shell structure
recorded from different independent measurements for gold, silver
and copper. We observe that the peak positions are very close to
each other for the three metals. There are variations for the gold
peaks indexed 6 and 9. It is possible that one peak is missing in
the histograms because of the supershell modulation of the peak
amplitudes, \cite{yanson00a}, as it will be explained later. The
standard deviation is quite low, showing that the stable diameters
can be reproduced very well in different measurements. We believe
that the small shifts that are observed come from variations in
the conductance due to back-scattering on defects near the contacts.\\

\subsection{Electronic shell effect theory}

\begin{figure}[b!]
\begin{center}
\includegraphics[width=8cm]{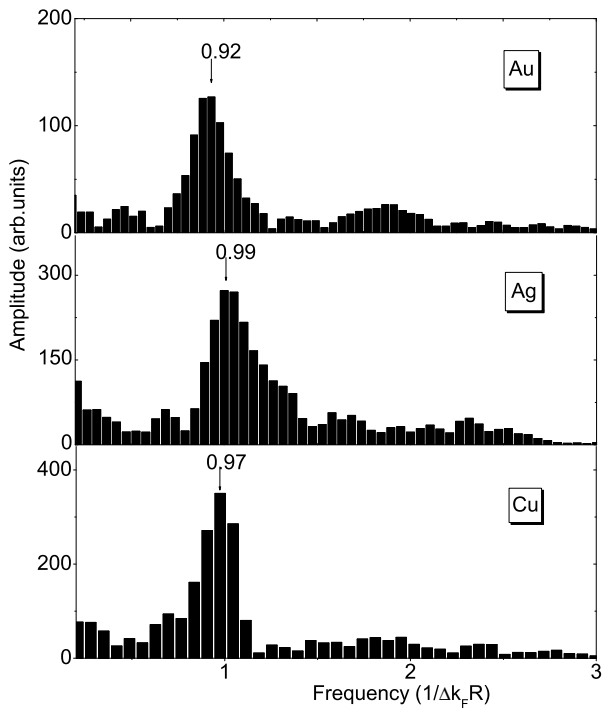}
\end{center}
\caption{Fourier transform of the conductance histograms showing
electronic shell effect for gold, silver and copper. The Fourier
spectra presents one main frequency that can be attributed to a
superposition of triangular and square orbit. The diametric orbit
expected at the frequency of $1/ \Delta k_{\rm F}R= 0.64$ might be
identified as the peak situated at about $1/ \Delta k_{\rm F}R=
0.68$ for silver and copper, being less pronounced for the latter.
For gold there is no clear peak around this value.} \label{FFT}
\end{figure}
The periodic pattern present in our histograms in Figs.
\ref{GAuESE} and \ref{GAgandCuESE} is due to minima in the
electronic free energy of the nanowire as function of elongation.
We compare our peak positions with the theoretically predicted
stable diameters reported by Ogando \textit{et al}.
\cite{Ogando02}. The theoretical model used is called stabilized
jellium model and considers the nanowire as an infinitely long
cylinder taking into account the average valence electron density
of the metal. With this assumption the energy oscillations as
function of radius are obtained, having minima due to shell
filling. These minima agree well with the experimentally obtained
stable diameters for the three metals in question, as we can see
in the insets of Figs. \ref{GAuESE} and \ref{GAgandCuESE}
(triangles) and in Fig.
\ref{GpeaksAgCuAu} (crosses).\\
The physical mechanism leading the magic series of diameters is
best illustrated using a semiclassical approach. The electron
moves classically in the circular cross section of the wire. The
stable diameters are determined by closed orbits inside the
cylindrical walls of the wire. The orbits that proved to have the
most significant contribution for alkali nanowires are the
diametric, triangular and square orbits \cite{yanson01t},
\cite{yanson00a}. The oscillating frequencies that result from
these orbits are $1/\Delta k_{\rm F}R= 0.64$ for diametric orbit
and $1/\Delta k_{\rm F}R= 0.83$, $1/\Delta k_{\rm F}R= 0.90$ for
triangular and square orbits. A beating effect known as supershell
effect appears due to the superposition of the diametric orbit
with the higher frequency orbits (triangular and square). In order
to separate the oscillating frequencies in the experimental
histograms we perform a Fourier transform. Since we are interested
in only the oscillatory part of the spectra in Figs. \ref{GAuESE}
and \ref{GAgandCuESE} we subtract a smooth background. The Fourier
transform for Au (Fig. \ref{FFT}, top) shows a broad peak centered
at a frequency of $1/ \Delta k_{\rm F}R= 0.92$. This value is
somewhat higher than what is expected from the superposition of
the triangular and square orbits. This deviation can be seen as
due to conductance lowering due to backscattering on defects in or
near the nanowire. This correction seems to be contact size
dependent, as seen by the fact that the conductance is lowered,
but the calculated radii of the contact are still linear with peak
index, as seen in the insets of Figs. \ref{GAuESE} and
\ref{GAgandCuESE}. Indeed the slope is somewhat lower than what
obtained from stabilized free electron model calculation
\cite{Ogando02} $\Delta k_{\rm F}R= 1.19 \pm 0.02$, and also by
comparison to the results on potassium and sodium nanowires
\cite{yanson01t}. However deviations of the same order have been
observed for lithium nanowires attributed to the same defect
scattering effect \cite{yanson01t}. In the case of silver and
copper the dominant peak in the Fourier transform is centered on
even higher frequencies $1/ \Delta k_{\rm F}R= 0.97$ and $1/
\Delta k_{\rm F}R= 0.99$.\\
The contribution of the diametric orbit seems to be less important
in the spectra for the noble metal nanowires as compared to the
alkali metals. We might identify a broad peak around $1/ \Delta
k_{\rm F}R= 0.68$ for silver and copper as being due to the
diametric orbit. Again, the frequency is somewhat higher than
predicted by the semiclassical model for a clean wire. For gold
there is no clear contribution of the diametric orbit. The
selective suppression of the diametric orbit may be explained in
terms of backscattering on surface roughness. For the circular
orbit the incoming electron wave is perpendicular to the surface
being therefore more probable to be diffusely scattered than in
the case of grazing incidence orbits. Another reason can be the
low resolution we have in the fourier transform because of the
limited number of peaks.\\
By applying a free electron model to noble metals one ignores the
non-spherical shape of their Fermi surfaces. The main sheets of
the bulk Fermi surface are connected by necks at Brillouin zone
boundaries along the [111] orientation. However, the contribution
of the necks to the oscillations in the density of states is
expected to be small since their wavelength is about six times
larger than the main Fermi wavelength, for gold and copper.
Filling of the states in the neck will have a period six times
larger than the one resulting from the states in the belly. The
silver Fermi surface has even smaller deviations, resulting in a
wavelength and a period of resulting oscillations of density of
states eight times higher than those in the belly. Moreover the
contribution of the states of the necks to the total density of
states is relatively small. Therefore, in a good approximation Au,
Ag,  and Cu may be considered free electron metals. Our assumption
is supported by electronic structure calculations for the quantum
modes in nanowires of Na and Cu \cite{opitz02}.

\subsection{Comparison to magic numbers of noble metal clusters}

We compare also the values for the preferred nanowire diameters
with the magic radii in clusters (circle symbols in Figs.
\ref{GAuESE} \ref{GAgandCuESE}, obtained from the number of atoms
in a cluster, N, as $\ k_{\rm F}R=1.919 N^{1/3}$ \cite{deheer93}).
We see that the agreement is very good. This is at first sight
unexpected due to the difference of symmetry, that is spherical in
the case of clusters and cylindrical for nanowires. However, we
first note that the gross features of distribution of zeros for
spherical and cylindrical Bessel functions are nearly identical
for not too large diameters. The difference between cylindrical
and spherical geometries is expressed mostly in the relative
weight of the various semiclassical orbits. For nanowires the
diametric orbit is expected to have a strong contribution in the
oscillation spectrum while for clusters it is negligible. Since we
have very little influence of the diametric orbit, possibly as a
result of surface roughness, we obtain about the same oscillation
period as for noble metal clusters.

\subsection{Atomic faceting}

At larger diameters, the surface energy becomes more important
than the free energy. The oscillation amplitudes of the electronic
free energy have a $1/R$ dependence \cite{yannouleas98} while the
ones for surface energy are roughly constant.  A crossover between
the two is experimentally observed by the change in the
oscillation period. We propose a model for nanowire faceting
starting from the crystalline order that we have in bulk: fcc for
all three noble metals. We assume that the nanowires form along
the [110] axis having a hexagonal cross section with four (111)
facets and two larger (100) ones (inset of
Fig.~\ref{GAuandAgASE}). The filling of each individual facet will
give a stable diameter. There has been another proposed cross
section of the nanowire with octagonal symmetry \cite{medina03}.
We have chosen the hexagonal cross section along the [110]
orientation supported by high resolution transmission electron
microscopy (HRTEM) observations \cite{rodrigues00},
\cite{rodrigues04}. These experiments provide evidence that the
bulk crystalline order survives in gold and copper atomic
contacts. The atomic arrangement of the nanowires obtained by
means of the Wulff construction reveal that the growth occurs
preferentialy along the crystalline directions [110], [111], and
[100], with the first one being more favorable for growing long
nanowires. Our model is further supported by Monte Carlo
simulations that confirm that for the process of thinning down of
a nanowire the [110] direction is a preferred orientation for
forming long and stable nanowires with a faceted structure
\cite{jagla01}. The expected periodicity of stable diameters is
$\Delta k_{\rm F}R=0.476$. This value is very close to the
experimentally observed periodicity for gold $\Delta k_{\rm
F}R=0.40$ and even closer for silver $\Delta k_{\rm F}R=0.46$.
Silver also has the largest number of atomic shell effect peaks,
as one can see in Fig.
\ref{GAuandAgASE}.\\
Previous results on copper nanowires in UHV-RT have been reported
by combining HRTEM and MCBJ \cite{rodrigues04}. From independent
imaging and conductance measurements of copper nanowires Gonzales
\textit{et al}. suggest that a stable pentagonal configuration
occurs having a conductance of 4.5 $G_{0}$.\\
In most of the cases (7 out of 10 measurements) our conductance
histograms for copper show a peak at 5 $G_{0}$, and very rarely at
lower values between 4 $G_{0}$ and 4.5 $G_{0}$. Similarly for
silver the peak position is close to $5G_{0}$. However for gold we
reproducibly see a distinct peak close to 4 $G_{0}$. This peak was
tentatively attributed to an quadrupolar distorted nanowire that
gold may have preference to form \cite{urbanTBP}. Such distortions
would be most likely when the surface tension is low. The surface
tension for gold lies in between that for Cu and Ag, which seems
to rule out this interpretation. We propose that the d-bonding
character for gold that also gives rise to the formation of atomic
chains \cite{smit01} may play a role for the smallest contacts.\\
Kondo {\it et al}. reported the formation and imaging of suspended
multi-shell helical gold nanowires with diameters ranging from 0.6
nm and length of 6 nm \cite{kondo00}. Such anomalous atomic
arrangements in nanowires, referred to as `weird wires', had been
predicted from model calculations by G{\" u}lseren {\it et al}.
\cite{gulseren98}. Recently the conductance of these structures
was calculated by first principle methods \cite{Ono05}. We compare
in Fig. \ref{GAuESE} the calculated values of the conductance for
the multi-shell helical wires with our peak positions. One can
observe that the period for the first few peaks is close to the
period of the calculated helical nanowire conductances, although
their values do not fully coincide. However, at higher
conductances the bars start to get closer together in contrast to
the peaks in the histogram. We do not exclude the formation of
helical nanowires, but we believe the peaks in the histogram are
due to shell effect considering the agreement with the
theoretically predicted period \cite{Ogando02}. The reason why
helical wires form in the experiment by Kondo {\it et al}. and not
in ours is likely to be attributed to the different experimental
methods for
forming the nanowires. \\

\section{Conclusion}

We have evidence that electronic shell filling influences the
formation and stability of all three noble metal nanowires: gold,
silver and copper. At larger diameters the atomic shell effect is
dominant and appears in gold and silver but was not observed in
copper. We observe that the shell structure is the most pronounced
in silver nanowires. Regarding the electronic shell structure the
Fourier spectrum reveals that the main contribution comes from the
superposition of triangular and square orbits. Free electron model
predictions of stable radii due to the shell effect agree well
with our results. Predicted values of conductance for gold
ellipticcally distorted nanowires agree with the experimental
peaks. Our stable diameters are in good agreement with the magic
diameters of noble metal clusters. Together with the results for
alkali metals \cite{yanson99,yanson01a,yanson01t,yanson01},  we
thus conclude that shell effects are generally observed for
monovalent metals. The effect is sufficiently robust that it can
be observed under ambient conditions for gold and silver.

\acknowledgments{We thank C.~A. Stafford for valuable discussions
and R. van Egmond for technical support. This work is part of the
research program of the ``Stichting FOM,'', and was further
supported by the European Commission TMR Network program DIENOW. }

\bibliographystyle{apsrev}
\bibliography{bib}

\begin{thebibliography}{30}
\expandafter\ifx\csname natexlab\endcsname\relax\def\natexlab#1{#1}\fi
\expandafter\ifx\csname bibnamefont\endcsname\relax
  \def\bibnamefont#1{#1}\fi
\expandafter\ifx\csname bibfnamefont\endcsname\relax
  \def\bibfnamefont#1{#1}\fi
\expandafter\ifx\csname citenamefont\endcsname\relax
  \def\citenamefont#1{#1}\fi
\expandafter\ifx\csname url\endcsname\relax
  \def\url#1{\texttt{#1}}\fi
\expandafter\ifx\csname urlprefix\endcsname\relax\def\urlprefix{URL }\fi
\providecommand{\bibinfo}[2]{#2}
\providecommand{\eprint}[2][]{\url{#2}}

\bibitem[{\citenamefont{Stafford et~al.}(1997)\citenamefont{Stafford,
  Baeriswyl, and B{\"u}rki}}]{stafford97}
\bibinfo{author}{\bibfnamefont{C.~A.} \bibnamefont{Stafford}},
  \bibinfo{author}{\bibfnamefont{D.}~\bibnamefont{Baeriswyl}},
  \bibnamefont{and}
  \bibinfo{author}{\bibfnamefont{J.}~\bibnamefont{B{\"u}rki}},
  \bibinfo{journal}{Phys. Rev. Lett.} \textbf{\bibinfo{volume}{79}},
  \bibinfo{pages}{2863} (\bibinfo{year}{1997}).

\bibitem[{\citenamefont{Yanson et~al.}(1999)\citenamefont{Yanson, Yanson, and
  van Ruitenbeek}}]{yanson99}
\bibinfo{author}{\bibfnamefont{A.~I.} \bibnamefont{Yanson}},
  \bibinfo{author}{\bibfnamefont{I.~K.} \bibnamefont{Yanson}},
  \bibnamefont{and} \bibinfo{author}{\bibfnamefont{J.~M.} \bibnamefont{van
  Ruitenbeek}}, \bibinfo{journal}{Nature} \textbf{\bibinfo{volume}{400}},
  \bibinfo{pages}{144} (\bibinfo{year}{1999}).

\bibitem[{\citenamefont{de~Heer}(1993)}]{deheer93}
\bibinfo{author}{\bibfnamefont{W.~A.} \bibnamefont{de~Heer}},
  \bibinfo{journal}{Rev. Mod. Phys.} \textbf{\bibinfo{volume}{65}},
  \bibinfo{pages}{611} (\bibinfo{year}{1993}).

\bibitem[{\citenamefont{Yanson et~al.}(2001{\natexlab{a}})\citenamefont{Yanson,
  Yanson, and van Ruitenbeek}}]{yanson01a}
\bibinfo{author}{\bibfnamefont{A.~I.} \bibnamefont{Yanson}},
  \bibinfo{author}{\bibfnamefont{I.~K.} \bibnamefont{Yanson}},
  \bibnamefont{and} \bibinfo{author}{\bibfnamefont{J.~M.} \bibnamefont{van
  Ruitenbeek}}, \bibinfo{journal}{Phys. Rev. Lett.}
  \textbf{\bibinfo{volume}{87}}, \bibinfo{pages}{216805}
  (\bibinfo{year}{2001}{\natexlab{a}}).

\bibitem[{\citenamefont{Mares et~al.}(2004)\citenamefont{Mares, Otte,
  Soukiassian, Smit, and van Ruitenbeek}}]{mares04}
\bibinfo{author}{\bibfnamefont{A.~I.} \bibnamefont{Mares}},
  \bibinfo{author}{\bibfnamefont{A.~F.} \bibnamefont{Otte}},
  \bibinfo{author}{\bibfnamefont{L.~G.} \bibnamefont{Soukiassian}},
  \bibinfo{author}{\bibfnamefont{R.~H.~M.} \bibnamefont{Smit}},
  \bibnamefont{and} \bibinfo{author}{\bibfnamefont{J.~M.} \bibnamefont{van
  Ruitenbeek}}, \bibinfo{journal}{Phys. Rev. B} \textbf{\bibinfo{volume}{70}},
  \bibinfo{pages}{073401} (\bibinfo{year}{2004}).

\bibitem[{\citenamefont{Hwang and Kang}(2003)}]{kwang03}
\bibinfo{author}{\bibfnamefont{H.~J.} \bibnamefont{Hwang}} \bibnamefont{and}
  \bibinfo{author}{\bibfnamefont{J.~W.} \bibnamefont{Kang}},
  \bibinfo{journal}{Surf. Sci.} \textbf{\bibinfo{volume}{532-535}},
  \bibinfo{pages}{536} (\bibinfo{year}{2003}).

\bibitem[{\citenamefont{B{\"u}rki et~al.}(2005)\citenamefont{B{\"u}rki,
  Stafford, and Stein}}]{burki05}
\bibinfo{author}{\bibfnamefont{J.}~\bibnamefont{B{\"u}rki}},
  \bibinfo{author}{\bibfnamefont{C.~A.} \bibnamefont{Stafford}},
  \bibnamefont{and} \bibinfo{author}{\bibfnamefont{D.~L.} \bibnamefont{Stein}}
  (\bibinfo{year}{2005}), \bibinfo{note}{preprint, cond-mat/0505221}.

\bibitem[{\citenamefont{Costa-Kr{\"a}mer}(1997)}]{costa97}
\bibinfo{author}{\bibfnamefont{J.~L.} \bibnamefont{Costa-Kr{\"a}mer}},
  \bibinfo{journal}{Phys. Rev. B} \textbf{\bibinfo{volume}{55}},
  \bibinfo{pages}{R4875} (\bibinfo{year}{1997}).

\bibitem[{\citenamefont{Yanson}(2001)}]{yanson01t}
\bibinfo{author}{\bibfnamefont{A.~I.} \bibnamefont{Yanson}}, Ph.D. thesis,
  \bibinfo{school}{Universiteit Leiden}, \bibinfo{address}{The Netherlands}
  (\bibinfo{year}{2001}).

\bibitem[{\citenamefont{Ono and Hirose}(2005)}]{Ono05}
\bibinfo{author}{\bibfnamefont{T.}~\bibnamefont{Ono}} \bibnamefont{and}
  \bibinfo{author}{\bibfnamefont{K.}~\bibnamefont{Hirose}},
  \bibinfo{journal}{Phys. Rev. Lett.} \textbf{\bibinfo{volume}{94}},
  \bibinfo{pages}{206806} (\bibinfo{year}{2005}).

\bibitem[{\citenamefont{Katakuse et~al.}(1985)\citenamefont{Katakuse, Ichihara,
  Fujita, Sakurai, and Matsuda}}]{katakuse85}
\bibinfo{author}{\bibfnamefont{I.}~\bibnamefont{Katakuse}},
  \bibinfo{author}{\bibfnamefont{T.}~\bibnamefont{Ichihara}},
  \bibinfo{author}{\bibfnamefont{Y.}~\bibnamefont{Fujita}},
  \bibinfo{author}{\bibfnamefont{T.}~\bibnamefont{Sakurai}}, \bibnamefont{and}
  \bibinfo{author}{\bibfnamefont{H.}~\bibnamefont{Matsuda}},
  \bibinfo{journal}{Int. J. Mass Spectrom. Ion Processes}
  \textbf{\bibinfo{volume}{67}}, \bibinfo{pages}{229} (\bibinfo{year}{1985}).

\bibitem[{\citenamefont{Ogando et~al.}(2002)\citenamefont{Ogando, Zabala, and
  Puska}}]{Ogando02}
\bibinfo{author}{\bibfnamefont{E.}~\bibnamefont{Ogando}},
  \bibinfo{author}{\bibfnamefont{N.}~\bibnamefont{Zabala}}, \bibnamefont{and}
  \bibinfo{author}{\bibfnamefont{M.}~\bibnamefont{Puska}},
  \bibinfo{journal}{Nanotechnology} \textbf{\bibinfo{volume}{13}},
  \bibinfo{pages}{363} (\bibinfo{year}{2002}).

\bibitem[{\citenamefont{Torres et~al.}(1994)\citenamefont{Torres, Pascual, and
  S{\'a}enz}}]{torres94}
\bibinfo{author}{\bibfnamefont{J.~A.} \bibnamefont{Torres}},
  \bibinfo{author}{\bibfnamefont{J.~I.} \bibnamefont{Pascual}},
  \bibnamefont{and} \bibinfo{author}{\bibfnamefont{J.~J.}
  \bibnamefont{S{\'a}enz}}, \bibinfo{journal}{Phys. Rev. B}
  \textbf{\bibinfo{volume}{49}}, \bibinfo{pages}{16581} (\bibinfo{year}{1994}).

\bibitem[{\citenamefont{H{\"o}ppler and Zwerger}(1998)}]{hoppler98}
\bibinfo{author}{\bibfnamefont{C.}~\bibnamefont{H{\"o}ppler}} \bibnamefont{and}
  \bibinfo{author}{\bibfnamefont{W.}~\bibnamefont{Zwerger}},
  \bibinfo{journal}{Phys. Rev. Lett.} \textbf{\bibinfo{volume}{80}},
  \bibinfo{pages}{1792} (\bibinfo{year}{1998}).

\bibitem[{\citenamefont{Yanson et~al.}(2001{\natexlab{b}})\citenamefont{Yanson,
  Yanson, and van Ruitenbeek}}]{yanson01}
\bibinfo{author}{\bibfnamefont{A.~I.} \bibnamefont{Yanson}},
  \bibinfo{author}{\bibfnamefont{I.~K.} \bibnamefont{Yanson}},
  \bibnamefont{and} \bibinfo{author}{\bibfnamefont{J.~M.} \bibnamefont{van
  Ruitenbeek}}, \bibinfo{journal}{Low Temp. Phys.}
  \textbf{\bibinfo{volume}{27}}, \bibinfo{pages}{1092}
  (\bibinfo{year}{2001}{\natexlab{b}}).

\bibitem[{\citenamefont{Scheer et~al.}(1998)\citenamefont{Scheer,
  {Agra\"{\i}t}, Cuevas, {Levy Yeyati}, Ludoph, Mart{\'\i}n-Rodero, {Rubio
  Bollinger}, van Ruitenbeek, and Urbina}}]{scheer98}
\bibinfo{author}{\bibfnamefont{E.}~\bibnamefont{Scheer}},
  \bibinfo{author}{\bibfnamefont{N.}~\bibnamefont{{Agra\"{\i}t}}},
  \bibinfo{author}{\bibfnamefont{J.~C.} \bibnamefont{Cuevas}},
  \bibinfo{author}{\bibfnamefont{A.}~\bibnamefont{{Levy Yeyati}}},
  \bibinfo{author}{\bibfnamefont{B.}~\bibnamefont{Ludoph}},
  \bibinfo{author}{\bibfnamefont{A.}~\bibnamefont{Mart{\'\i}n-Rodero}},
  \bibinfo{author}{\bibfnamefont{G.}~\bibnamefont{{Rubio Bollinger}}},
  \bibinfo{author}{\bibfnamefont{J.~M.} \bibnamefont{van Ruitenbeek}},
  \bibnamefont{and} \bibinfo{author}{\bibfnamefont{C.}~\bibnamefont{Urbina}},
  \bibinfo{journal}{Nature} \textbf{\bibinfo{volume}{394}},
  \bibinfo{pages}{154} (\bibinfo{year}{1998}).

\bibitem[{\citenamefont{Hansen et~al.}(2000)\citenamefont{Hansen, Nielsen,
  Brandbyge, L{\ae}gsgaard, Stensgaard, and Besenbacher}}]{hansen00}
\bibinfo{author}{\bibfnamefont{K.}~\bibnamefont{Hansen}},
  \bibinfo{author}{\bibfnamefont{S.~K.} \bibnamefont{Nielsen}},
  \bibinfo{author}{\bibfnamefont{M.}~\bibnamefont{Brandbyge}},
  \bibinfo{author}{\bibfnamefont{E.}~\bibnamefont{L{\ae}gsgaard}},
  \bibinfo{author}{\bibfnamefont{I.}~\bibnamefont{Stensgaard}},
  \bibnamefont{and}
  \bibinfo{author}{\bibfnamefont{F.}~\bibnamefont{Besenbacher}},
  \bibinfo{journal}{Appl. Phys. Lett.} \textbf{\bibinfo{volume}{77}},
  \bibinfo{pages}{708} (\bibinfo{year}{2000}).

\bibitem[{\citenamefont{Hansen et~al.}(1997)\citenamefont{Hansen, Laegsgaard,
  Stensgaard, and Besenbacher}}]{hansen97}
\bibinfo{author}{\bibfnamefont{K.}~\bibnamefont{Hansen}},
  \bibinfo{author}{\bibfnamefont{E.}~\bibnamefont{Laegsgaard}},
  \bibinfo{author}{\bibfnamefont{I.}~\bibnamefont{Stensgaard}},
  \bibnamefont{and}
  \bibinfo{author}{\bibfnamefont{F.}~\bibnamefont{Besenbacher}},
  \bibinfo{journal}{Phys. Rev. B} \textbf{\bibinfo{volume}{56}},
  \bibinfo{pages}{2208} (\bibinfo{year}{1997}).

\bibitem[{\citenamefont{Krans et~al.}(1993)\citenamefont{Krans, Muller, Yanson,
  Govaert, Hesper, and van Ruitenbeek}}]{krans93}
\bibinfo{author}{\bibfnamefont{J.~M.} \bibnamefont{Krans}},
  \bibinfo{author}{\bibfnamefont{C.~J.} \bibnamefont{Muller}},
  \bibinfo{author}{\bibfnamefont{I.~K.} \bibnamefont{Yanson}},
  \bibinfo{author}{\bibfnamefont{T.~C.~M.} \bibnamefont{Govaert}},
  \bibinfo{author}{\bibfnamefont{R.}~\bibnamefont{Hesper}}, \bibnamefont{and}
  \bibinfo{author}{\bibfnamefont{J.~M.} \bibnamefont{van Ruitenbeek}},
  \bibinfo{journal}{Phys. Rev. B} \textbf{\bibinfo{volume}{48}},
  \bibinfo{pages}{14721} (\bibinfo{year}{1993}).

\bibitem[{\citenamefont{Yanson et~al.}(2000)\citenamefont{Yanson, Yanson, and
  van Ruitenbeek}}]{yanson00a}
\bibinfo{author}{\bibfnamefont{A.~I.} \bibnamefont{Yanson}},
  \bibinfo{author}{\bibfnamefont{I.~K.} \bibnamefont{Yanson}},
  \bibnamefont{and} \bibinfo{author}{\bibfnamefont{J.~M.} \bibnamefont{van
  Ruitenbeek}}, \bibinfo{journal}{Phys. Rev. Lett.}
  \textbf{\bibinfo{volume}{84}}, \bibinfo{pages}{5832} (\bibinfo{year}{2000}).

\bibitem[{\citenamefont{Opitz et~al.}(2002)\citenamefont{Opitz, Zahn, and
  Mertig}}]{opitz02}
\bibinfo{author}{\bibfnamefont{J.}~\bibnamefont{Opitz}},
  \bibinfo{author}{\bibfnamefont{P.}~\bibnamefont{Zahn}}, \bibnamefont{and}
  \bibinfo{author}{\bibfnamefont{I.}~\bibnamefont{Mertig}},
  \bibinfo{journal}{Phys. Rev. B} \textbf{\bibinfo{volume}{66}},
  \bibinfo{pages}{245417} (\bibinfo{year}{2002}).

\bibitem[{\citenamefont{Yannouleas et~al.}(1998)\citenamefont{Yannouleas,
  Bogachek, and Landman}}]{yannouleas98}
\bibinfo{author}{\bibfnamefont{C.}~\bibnamefont{Yannouleas}},
  \bibinfo{author}{\bibfnamefont{E.~N.} \bibnamefont{Bogachek}},
  \bibnamefont{and} \bibinfo{author}{\bibfnamefont{U.}~\bibnamefont{Landman}},
  \bibinfo{journal}{Phys. Rev. B} \textbf{\bibinfo{volume}{57}},
  \bibinfo{pages}{4872} (\bibinfo{year}{1998}).

\bibitem[{\citenamefont{Medina et~al.}(2004)\citenamefont{Medina, D{\'i}az,
  L{\'e}on, Guerrero, Hasmy, Serena, and Costa-Kr{\"a}mer}}]{medina03}
\bibinfo{author}{\bibfnamefont{E.}~\bibnamefont{Medina}},
  \bibinfo{author}{\bibfnamefont{M.}~\bibnamefont{D{\'i}az}},
  \bibinfo{author}{\bibfnamefont{N.}~\bibnamefont{L{\'e}on}},
  \bibinfo{author}{\bibfnamefont{C.}~\bibnamefont{Guerrero}},
  \bibinfo{author}{\bibfnamefont{A.}~\bibnamefont{Hasmy}},
  \bibinfo{author}{\bibfnamefont{P.~A.} \bibnamefont{Serena}},
  \bibnamefont{and} \bibinfo{author}{\bibfnamefont{J.~L.}
  \bibnamefont{Costa-Kr{\"a}mer}}, \bibinfo{journal}{Phys. Rev. Lett.}
  \textbf{\bibinfo{volume}{93}}, \bibinfo{pages}{186403}
  (\bibinfo{year}{2004}).

\bibitem[{\citenamefont{Rodrigues et~al.}(2000)\citenamefont{Rodrigues, Fuhrer,
  and Ugarte}}]{rodrigues00}
\bibinfo{author}{\bibfnamefont{V.}~\bibnamefont{Rodrigues}},
  \bibinfo{author}{\bibfnamefont{T.}~\bibnamefont{Fuhrer}}, \bibnamefont{and}
  \bibinfo{author}{\bibfnamefont{D.}~\bibnamefont{Ugarte}},
  \bibinfo{journal}{Phys. Rev. Lett.} \textbf{\bibinfo{volume}{85}},
  \bibinfo{pages}{4124} (\bibinfo{year}{2000}).

\bibitem[{\citenamefont{Gonzalez et~al.}(2004)\citenamefont{Gonzalez,
  Rodrigues, Bettini, Rego, Rocha, Coura, Dantas, Sato, Galvao, and
  Ugarte}}]{rodrigues04}
\bibinfo{author}{\bibfnamefont{J.~C.} \bibnamefont{Gonzalez}},
  \bibinfo{author}{\bibfnamefont{V.}~\bibnamefont{Rodrigues}},
  \bibinfo{author}{\bibfnamefont{J.}~\bibnamefont{Bettini}},
  \bibinfo{author}{\bibfnamefont{L.~G.~C.} \bibnamefont{Rego}},
  \bibinfo{author}{\bibfnamefont{A.~R.} \bibnamefont{Rocha}},
  \bibinfo{author}{\bibfnamefont{P.~Z.} \bibnamefont{Coura}},
  \bibinfo{author}{\bibfnamefont{S.~O.} \bibnamefont{Dantas}},
  \bibinfo{author}{\bibfnamefont{F.}~\bibnamefont{Sato}},
  \bibinfo{author}{\bibfnamefont{D.~S.} \bibnamefont{Galvao}},
  \bibnamefont{and} \bibinfo{author}{\bibfnamefont{D.}~\bibnamefont{Ugarte}},
  \bibinfo{journal}{Phys. Rev. Lett.} \textbf{\bibinfo{volume}{93}},
  \bibinfo{pages}{126103} (\bibinfo{year}{2004}).

\bibitem[{\citenamefont{Jagla and Tosatti}(2001)}]{jagla01}
\bibinfo{author}{\bibfnamefont{E.~A.} \bibnamefont{Jagla}} \bibnamefont{and}
  \bibinfo{author}{\bibfnamefont{E.}~\bibnamefont{Tosatti}},
  \bibinfo{journal}{Phys. Rev. B} \textbf{\bibinfo{volume}{64}},
  \bibinfo{pages}{205412} (\bibinfo{year}{2001}).

\bibitem[{\citenamefont{Urban et~al.}(2004)\citenamefont{Urban, B{\"u}rki,
  Zhang, Stafford, and Grabert}}]{urbanTBP}
\bibinfo{author}{\bibfnamefont{D.~F.} \bibnamefont{Urban}},
  \bibinfo{author}{\bibfnamefont{J.}~\bibnamefont{B{\"u}rki}},
  \bibinfo{author}{\bibfnamefont{C.-H.} \bibnamefont{Zhang}},
  \bibinfo{author}{\bibfnamefont{C.~A.} \bibnamefont{Stafford}},
  \bibnamefont{and} \bibinfo{author}{\bibfnamefont{H.}~\bibnamefont{Grabert}},
  \bibinfo{journal}{Phys. Rev. Lett.} \textbf{\bibinfo{volume}{93}},
  \bibinfo{pages}{186403} (\bibinfo{year}{2004}).

\bibitem[{\citenamefont{Smit et~al.}(2001)\citenamefont{Smit, Untiedt, Yanson,
  and van Ruitenbeek}}]{smit01}
\bibinfo{author}{\bibfnamefont{R.~H.~M.} \bibnamefont{Smit}},
  \bibinfo{author}{\bibfnamefont{C.}~\bibnamefont{Untiedt}},
  \bibinfo{author}{\bibfnamefont{A.~I.} \bibnamefont{Yanson}},
  \bibnamefont{and} \bibinfo{author}{\bibfnamefont{J.~M.} \bibnamefont{van
  Ruitenbeek}}, \bibinfo{journal}{Phys. Rev. Lett.}
  \textbf{\bibinfo{volume}{87}}, \bibinfo{pages}{266102}
  (\bibinfo{year}{2001}).

\bibitem[{\citenamefont{Kondo and Takayanagi}(2000)}]{kondo00}
\bibinfo{author}{\bibfnamefont{Y.}~\bibnamefont{Kondo}} \bibnamefont{and}
  \bibinfo{author}{\bibfnamefont{K.}~\bibnamefont{Takayanagi}},
  \bibinfo{journal}{Science} \textbf{\bibinfo{volume}{289}},
  \bibinfo{pages}{606} (\bibinfo{year}{2000}).

\bibitem[{\citenamefont{G{\"u}lseren et~al.}(1998)\citenamefont{G{\"u}lseren,
  Ercolessi, and Tosatti}}]{gulseren98}
\bibinfo{author}{\bibfnamefont{O.}~\bibnamefont{G{\"u}lseren}},
  \bibinfo{author}{\bibfnamefont{F.}~\bibnamefont{Ercolessi}},
  \bibnamefont{and} \bibinfo{author}{\bibfnamefont{E.}~\bibnamefont{Tosatti}},
  \bibinfo{journal}{Phys. Rev. Lett.} \textbf{\bibinfo{volume}{80}},
  \bibinfo{pages}{3775} (\bibinfo{year}{1998}).

\end{thebibliography}

\end{document}